\RequirePackage{fix-cm}  % to remove annoying warnings
\documentclass[aps,prl,showpacs,amsmath,amssymb,amsfonts,superscriptaddress,twocolumn,lengthcheck]{revtex4-1}
\usepackage{graphicx}
\usepackage{subfigure}
\usepackage{amsthm}
\usepackage{verbatim}
\usepackage{dcolumn}% Align table columns on decimal point
\usepackage{bm}% bold math
\usepackage{epsf}
\usepackage{color}
\usepackage[colorlinks=true,citecolor=blue,linkcolor=blue,urlcolor=blue]{hyperref}%
\newcommand{\bra}[1]{\left\langle #1\right|}
\newcommand{\ket}[1]{\left|#1\right\rangle}
\newcommand{\braket}[2]{\left\langle #1|#2\right\rangle}

\newcommand{\tr}[1]{\mathrm{tr}\left\{#1\right\}}

\newcommand{\id}{\mathbb{I}}

\newcommand{\bla}{bla\\bla\\bla\\bla\\bla}

\newcommand{\mc}[1]{\mathcal{#1}}

\begin{document}
	%
	%\title{Decoherence in $\mathcal{P}\mathcal{T}$-symmetric quantum systems}
	\title{$\mathcal{P}\mathcal{T}$-symmetric slowing-down of decoherence}
	\author{Bart\l{}omiej Gardas}
	\affiliation{Theoretical Division, Los Alamos National Laboratory, Los Alamos, NM 87545, USA}
	\affiliation{Institute of Physics, University of Silesia, 40-007 Katowice, Poland}
	\author{Sebastian Deffner}
	\affiliation{Theoretical Division, Los Alamos National Laboratory, Los Alamos, NM 87545, USA}
	\affiliation{Department of Physics, University of Maryland Baltimore County, Baltimore, MD 21250, USA}
	\affiliation{Center for Nonlinear Studies, Los Alamos National Laboratory, Los Alamos, NM 87545, USA}
	\author{Avadh Saxena}
	\affiliation{Theoretical Division, Los Alamos National Laboratory, Los Alamos, NM 87545, USA}
	\affiliation{Center for Nonlinear Studies, Los Alamos National Laboratory, Los Alamos, NM 87545, USA}
	\date{\today}
	
	\begin{abstract}		
	We investigate $\mc{P}\mc{T}$-symmetric quantum systems ultra-weakly coupled to an environment. 
	We find that such open systems evolve under $\mc{P}\mc{T}$-symmetric, purely dephasing and unital dynamics.
	The dynamical map describing the evolution is then determined explicitly using a quantum canonical transformation. 
	Furthermore, we provide an explanation of why $\mc{P}\mc{T}$-symmetric dephasing type interactions lead to 
	\emph{critical slowing down of decoherence}. This effect is further exemplified with an experimentally relevant 
	system -- a $\mc{P}\mc{T}$-symmetric qubit easily realizable, \emph{e.g.}, in optical or microcavity experiments.
	\end{abstract}
	
	 % Quantum mechanics, Foundations of quantum mechanics; measurement theory, Formalism
	\maketitle
	
\paragraph{Introduction.}
Symmetry is one of the most important and profound concepts in physics~\cite{Weyl52,Gross96} which explains the \emph{modus operandi} 
of many complex physical and biological systems~\cite{Breier16}. It expresses how systems remain unaffected by perturbations~\cite{Peres95}. 
Therefore, a violation of symmetry (or its breakdown~\cite{Zhang13}) constitutes an irreplaceable source of valuable information regarding
properties of physical systems~\cite{Higgs64,Ge14,Zhu16}. There is an abundance of useful transformations providing necessary ingredients to
understand and investigate quantum systems. Among them, there are two of special physical significance: the time reversal operation 
$\mc{T}$~\cite{Crooks08} and parity -- a mirror-reflection symmetry -- $\mc{P}$~\cite{Zeng05}. These two transformations are both hermitian 
and independent of each other, \emph{i.e.}, $[\mc{P},\mc{T}]=0$. Systems that are invariant under the joint $\mc{P}\mc{T}$ operation
are called $\mc{P}\mc{T}$-symmetric~\cite{bender_1}. Such effectively open systems exhibit dynamics with balanced loss and gain~\cite{Jing14,Sounas16}. Recent results have proven to be of great theoretical~\cite{Gardas16,Kartashov15,Hang13,Bender1998} and experimental~\cite{Lawrence14,optic,optic_2,gao} importance, and $\mc{P}\mc{T}$-symmetric quantum systems have been realized in many different setups, such as optical~\cite{Guo09}, 
optomechanical~\cite{Lu15} or microcavity based experiments~\cite{Peng14,Jing14}.

Contemporary studies have revealed that important (non)equilibrium properties and thermodynamic relations also hold for $\mc{P}\mc{T}$-symmetric 
quantum systems; \emph{e.g.}, the Carnot theorem~\cite{Gardas16,Jiang15} and the Jarzynski equality~\cite{Jarzynski97,Deffner2015a}. Nevertheless, to further 
advance our understanding of $\mc{P}\mc{T}$-symmetric quantum systems the next natural step is to understand decoherence~\cite{Zurek03}. This is particularly
important when one wants to store and process information in quantum systems~\cite{Mohseni03,Ladd10,Pellizzari95}. 

A comprehensive description of the system's dynamics requires tracing out the environmental degrees of freedom. Unfortunately, except for a 
few analytically solvable models, finding such reduced dynamics $\varrho_{\text{S}}(t)$ has proven to be extremely complicated, often impossible, 
even for hermitian systems~\cite{Breuer02}. Recently, it has been shown that all $\mc{P}\mc{T}$-symmetric quantum systems that admit real spectrum 
can be represented in a physically equivalent way by hermitian Hamiltonians~\cite{Gardas16b}. One would therefore expect them to be influenced by 
decoherence in a similar manner. In this Letter, however, we demonstrate features that are unique to $\mc{P}\mc{T}$-symmetric systems, resulting 
from the way they interact with their environment. In particular, we investigate a $\mc{P}\mc{T}$-symmetric quantum system coupled ultra-weakly to a 
hermitian environment~\cite{Spohn07}. Our motivation is twofold: First, very weak coupling guarantees that no heat is exchanged between the system 
and environment~\cite{Averin16,Thoma16}. This leads to a phenomenon known as pure decoherence or dephasing~\cite{Cummings16,Chesi16}. Only 
quantum information is allowed to enter or leave the system so that any effect caused solely by decoherence can be quantified easily. 
Finally, following the Ockham's razor principle~\cite{Sober15}, hermiticity of the environment is assumed for the sake of simplicity and transparency 
of our description. 

Under these assumptions, we find that  such open systems evolve under $\mc{P}\mc{T}$-symmetric, purely dephasing \emph{and} 
unital dynamics. The dynamical map describing the evolution is then determined explicitly using a quantum canonical transformation. Therefore, 
as an immediate consequence of dephasing \emph{and} unital dynamics we find the validity of the Jarzynski equality~\cite{Rastegin14}. Furthermore,
we explain how a $\mc{P}\mc{T}$-symmetric dephasing channel leads to \emph{critical slowing down of decoherence}. This effect is exemplified using 
an experimentally relevant example -- a $\mc{P}\mc{T}$-symmetric qubit. Such a two-level system can be realized \emph{e.g.} in optics~\cite{optic} 
or in a microcavity~\cite{gao}. In particular, in the development of practical architectures for quantum computer systems with minimal or suppressed
decoherence are appealing~\cite{Lidar98,Johnson11}. We will see that $\mc{P}\mc{T}$-symmetric qubits are thus significantly better suited than standard, 
hermitian qubits~\cite{Lanting14}.

\paragraph{Pure decoherence in $\mathcal{P}\mathcal{T}$-symmetric quantum systems.}
	Consider a $\mathcal{P}\mathcal{T}$-symmetric quantum system $S$ interacting with its environment, $B$. 
	The composite system $S+B$ can be described by the following Hamiltonian
	\begin{equation}
		\label{htotal}
		  H = H_{\text{S}} \otimes \id_{\text{B}} + \id_{\text{S}} \otimes H_{\text{B}} + H_{\text{I}},
		  %H = H_{\text{S}} + H_{\text{B}} + H_{\text{I}}
	\end{equation}
	where $H_{\text{S}}$ and $H_{\text{B}}$ are the Hamiltonians of the system and the environment respectively, and $H_{\text{I}}$
	describes the interaction between them. In the following, we assume the usual form of the interaction: $H_{\text{I}} = V_{\text{S}}
	\otimes V_{\text{B}}$, where both $H_{\text{B}}$ and $V_{\text{B}}$ are hermitian yet $V_{\text{S}}$ is $\mc{P}\mc{T}$-symmetric.
	Typical examples include $\mc{P}\mc{T}$-symmetric resonators coupled weakly to the rest of the (hermitian) Universe~\cite{Phang15}. 
	A particularly interesting example arises when $V_{\text{S}} = g(H_{\text{S}})$, where $g$ is an arbitrary function. 
	Since $[H_{\text{S}},H_{\text{I}}]=0$, there is no energy exchange between the system and its environment; \emph{i.e.}, 
	$\langle H_{\text{S}}\rangle$ remains constant during the evolution. Therefore, any effect of the environment on the system leads to 
	pure decoherence~\cite{Alicki04}. Without any loss of generality we further assume that $g(H_{\text{S}})=H_{\text{S}}$.
	
	Henceforth, we focus on $\mathcal{P}\mathcal{T}$-symmetric quantum systems and show how to construct their reduced dynamics in the 
	presence of pure decoherence. To this end, we notice that if the spectrum of the system is real a hermitian transformation $T$ such
	that $h_{\text{S}} = T H_{\text{S}} T^{-1}$ is hermitian can always be found~\cite{Gardas16}. Moreover, since $h_{\text{S}}$ is hermitian
	we also have $H_{\text{S}}^{\dagger} = T^2 H_{\text{S}} T^{-2}$ which will be crucial for our analysis. We will prove this shortly.
	Here, we only note that in order to change hermiticity such a transformation cannot be unitary. However, $T$ preserves (canonical) 
	commutation relations (\emph{e.g.} between $x$ and $p$: $[x,p]=i\hbar$) and therefore will be regarded as a quantum canonical 
	transformation~\cite{Anderson94,Lee95}. More importantly, canonical transformations do \emph{not} change expectation values
	of observables: $\langle O_{\text{S}}\rangle_{\mathcal{P}\mathcal{T}} =\langle o_{\text{S}}\rangle_{H}$ where $o_{\text{S}}=TO_{\text{S}}T^{-1}$.
	
	Now, applying the canonical transformation $T$ to the Hamiltonian~(\ref{htotal}) yields
	\begin{equation}
		\label{ntotal}
		    h = h_{\text{S}} \otimes \id_{\text{B}} + \id_{\text{S}} \otimes H_{\text{B}} + h_{\text{I}},
		    \quad h_{\text{I}}=T H_{\text{I}} T^{-1},
	\end{equation}		
	where $T$ acts nontrivially only on the system of interest. Since the two systems are now hermitian their composed 
	dynamics is described by the Liouville-von Neumann equation of motion, $i\dot{\varrho}(t)=[h,\varrho(t)]$, whose 
	unique solution can be written as~\cite{Breuer02}
	\begin{equation}
		\varrho(t) \rightarrow U(t)\varrho(0)U(t)^{\dagger}, \quad U(t) = \exp(-iht).
	\end{equation}
	At any given time $t$, the reduced system's dynamics is determined by tracing out the environmental degrees of 
	freedom (see \emph{e.g.}~\cite{Alicki07}). Thus, one can write
	\begin{equation}
	   \label{ptr}
		\varrho_{\text{S}}(t) = \text{tr}_{\text{B}}\{U(t) \varrho_{\text{S}}(0) \otimes \Omega_{\text{B}} U(t)^{\dagger}\},
	\end{equation}
	where $\Omega_{\text{B}}$ is the initial state of the environment and $\text{tr}_{\text{B}}\{\cdot\}$ denotes the partial trace~\cite{Alicki07}.
    Note that the two systems are uncorrelated at $t=0$~\cite{Ringbauer15}. This requirement is crucial for the map $\Phi$: $\varrho_{\text{S}}(t)=
    	\Phi[\varrho_{\text{S}}(0)]$ to be well defined~\cite{Buzek01,*Buzek01a}. However, this is not difficult to fulfill experimentally~\cite{Chen16}.
			
	Since $\Omega_{\text{B}}$ is a density operator, it can be expressed as $\Omega_{\text{B}}=\sum_{\alpha}p_{\alpha}\ket{\alpha}\bra{\alpha}$, 
	where $p_\alpha$ denotes the probability of finding the environment in state $\ket{\alpha}$. As a result, the reduced dynamics~(\ref{ptr}) 
	can be rewritten using the so called operator-sum representation~\cite{Schlosshauer05}: 
	\begin{equation}
		\label{kraus}
			\varrho_{\text{S}}(t) = \textstyle{\sum_i} K_i(t)  \varrho_{\text{S}}(0) K_i(t)^{\dagger},
	\end{equation}
	where the Kraus operators $K_i(t)=\sqrt{p_{\alpha}}\bra{\beta}U(t)\ket{\alpha}$ satisfy $\sum_i K_i(t)^{\dagger} K_i(t)=\id_{\text{S}}$. 
	To simplify notation we have combined the two indices $\alpha$, $\beta$ into  $i$. Moreover, Eq.~(\ref{kraus}) defines a 
	\emph{unital} map, \emph{i.e.} $\Phi[\id_{\text{S}}]=\id_{\text{S}} $~\cite{Rastegin14}. Indeed, since $h_{\text{S}}$ commutes with $h$ 
	we also have $[K_i(t),K_i(t)^{\dagger}]=0$~\footnote{$K_i$, $K_i^{\dagger}$ commute because $K_i=f_i(h_{\text{S}})$ and $K_i^{\dagger}=g_i(h_{\text{S}})$.}.

	The operator-sum representation in Eq.~(\ref{kraus}) provides the most general description of decoherence and dissipation for hermitian
	quantum systems, which results from the interaction with the environment. It is often referred to as a quantum channel, \emph{i.e.}, a map
	that is completely positive and trace preserving - CPTP~\cite{Nielsen11}. When there is only one Kraus operator the evolution is 
	unitary~\footnote{This is due to the normalization, $\sum_i K_i(t)^{\dagger} K_i(t)=\id_{\text{S}}$.}. 
	Multiplying Eq.~(\ref{kraus}) from both sides by $T^{-1}$ and $T$, respectively, yields
    \begin{equation}
       \label{nkraus}
     	\rho_{\text{S}}(t) = T^{-1}\varrho_{\text{S}}(t)T = \textstyle{\sum_i} L_i(t) \rho_{\text{S}}(0) R_i(t),
    \end{equation}				
	where the left, $L_i(t)$, and right, $R_i(t)$, Kraus operators read
	\begin{equation}
		\label{LR}
		L_i(t) = T^{-1} K_i(t) T, \quad R_i(t) = T^{-1} K_i(t)^{\dagger} T.
	\end{equation}
    We see immediately that they fulfill $\sum_i L_i(t)R_i(t) = \id_{\text{S}}$. The last equality assures that the $\mathcal{P}\mathcal{T}$-CPTP 
	map~(\ref{LR}) is unital as well~\cite{Rastegin13,Kafri12,Albash14}. Therefore, $\mc{P}\mc{T}$-symmetric, purely dephasing and unital dynamics
	preserve the Jarzynski equality~\footnote{Technically, for the Jarzynski equality to apply one needs time dependence. 
	Time-dependent systems, however, can be treated with techniques described in~\cite{Gardas16} or~\cite{Deffner2015a}.}. Similar conclusions have
	been drawn recently for $\mathcal{P}\mathcal{T}$-symmetric Schr{\"o}dinger dynamics~\cite{Deffner2015a}. Note, when the dynamics is unitary then 			
	$L(t)=U(t)$ and $R(t)=U(-t)$, where $U(t)$ satisfies the Schr{\"o}dinger equation. We emphasize, however, that 
	$U(t)^{\dagger}\not=U(-t)$.
	
	In summary, Eq.~(\ref{nkraus}) provides the most general description of open $\mathcal{P}\mathcal{T}$-symmetric quantum systems. This is 
	our main result. To this end, we followed the following recipe: First, one transforms the $\mathcal{P}\mathcal{T}$-symmetric Hamiltonian into
	its hermitian representation using a quantum canonical transformation. Next, after solving the corresponding equation of motion, the inverse
	map is applied to obtain the final solution~\footnote{Note, obtaining $\rho_{\text{S}}(t)$ is not necessary to compute expectation values as 
	these remain the same in both representations.}. Our approach is generic and can be applied to \emph{e.g.} Lindblad master equations~\cite{Dast14,Prosen12} or 
	quantum Brownian motion~\cite{Lampo16,Hanggi09}. Also, our strategy is not restricted just to Markovian dynamics~\cite{Zhang12}. 
	However, for the present purposes we have chosen a model without heat exchange between system and environment~\cite{Alicki04}. 
	\paragraph{Canonical transformation.}
	As we have seen, to obtain the reduced dynamics for an open $\mathcal{P}\mathcal{T}$-symmetric system one needs to construct a canonical 
	transformation $T$ that restores hermiticity~\cite{Gardas16}. To this end, we assume that all energies of $H_{\text{S}}$ are real and 
	experimentally accessible. For the sake of simplicity, we also assume that the spectrum of $H_{\text{S}}$ is discrete and non-degenerate. 
	Therefore, there exists a basis $\ket{E_n}$ in which all energies $E_n$ can be measured. Hence, 
	$V H_{\text{S}} V^{-1} = \sum_n E_n \ket{E_n}\bra{E_n}$, where $\braket{E_n}{E_m}=\delta_{nm}$ and all energies $E_n$ are real. 
	Now, the canonical transformation $T$ can be calculated as $T=\sqrt{V^{\dagger}V}$. Note, since $H_{\text{S}}$ is not hermitian 
	$V$ is not unitary (\emph{i.e.}, $V^{\dagger}\not=V^{-1}$). To show this elegant and simple result we first notice that $H_{\text{S}}$ 
	 can also be rewritten as~\footnote{This simple result is well-know in linear algebra.}
	\begin{equation}
		\label{bi}
		    H_{\text{S}} = \textstyle{\sum_n} E_n \ket{\psi_n}\bra{\phi_n}.
	\end{equation}
	The new eigenstates $\ket{\psi_n}=V^{-1}\ket{E_n}$ and $\bra{\phi_n}=\bra{E_n}V$ form a biorthonormal basis~\cite{basic,Mostafazadeh02}. 
	That is to say, the following orthogonality and completeness relations hold: $\braket{\psi_n}{\phi_n}=\delta_{nm}$ and $\sum_n 
	\ket{\psi_n}\bra{\phi_n} = \id_{\text{S}}$. Biorthonormality also means that $\ket{\psi_n}$, $\ket{\phi_n}$ are the left and right eigenstates 
	of $H_{\text{S}}$, respectively. The corresponding eigenenergy reads $E_n$. Since $H_{\text{S}}$ is $\mathcal{P}\mathcal{T}$-symmetric, 
	it follows that~\cite{Weigert03} 
	\begin{equation}
		\label{pt}
			\mathcal{P}H_{\text{S}}\mathcal{P}=\mathcal{T}H_{\text{S}}\mathcal{T} = H_{\text{S}}^{\dagger}.
	\end{equation}	
	From the last equation we have $\mathcal{P}\ket{\psi_n}=e^{i\theta_n}\ket{\phi_n}$, where $\theta_n=0$, $\pi$. 
	Now, $T$ can be decomposed as $T^2=\mathcal{P}\mathcal{C}$, where the charge conjugation $\mathcal{C}$ reads~\cite{Weigert03}
	\begin{equation}
	   \label{C}
	       \mathcal{C} = \textstyle{\sum_n} \ket{\psi_n}\bra{\psi_n}\mathcal{P}
	       \quad \text{(note} \quad \mathcal{C}^2=\id_{\text{S}} \text{)}.
	\end{equation}
	By construction, the charge conjugation commutes with the system Hamiltonian $H_{\text{S}}$ and thus from Eq.~(\ref{pt})
	it follows immediately that $T^2\,H_{\text{S}}\,T^{-2} = H_{\text{S}}^{\dagger}$~\footnote{Note that $T^2=\sum_n \unexpanded{\ket{\phi_n}\bra{\phi_n}}$ and $T^{-2}=\sum_n \unexpanded{\ket{\psi_n}\bra{\psi_n}}$}. Finally,
	\begin{equation}
	   \label{her}
		  %H_{\text{S}} \left(\mathcal{C}\mathcal{P}\right)^{-1} \left(\mathcal{C}\mathcal{P}\right)^{1/2} = h_{\text{S}}.
		  h_{\text{S}}^{\dagger} = T^{-1} \left(T^2\,H_{\text{S}}\,T^{-2}\right) T = h_{\text{S}}.
	\end{equation}
	In conclusion, the canonical map $T$ indeed transforms a $\mathcal{P}\mathcal{T}$-symmetric Hamiltonian $H_{\text{S}}$ into
	a hermitian one, $h_{\text{S}}$. The main results~(\ref{nkraus})-(\ref{LR}) hold for all $\mathcal{P}\mathcal{T}$-symmetric 
	quantum systems that admit real spectra~\footnote{This result is a special case of a general theory 
	of pseudo-hermitian quantum systems~\cite{Gardas16}.}.

	\paragraph{Critical slowing down of  decoherence.}
	The remainder of our work is dedicated to studying an experimentally relevant example~\cite{gao}.
	Consider a $\mathcal{P}\mathcal{T}$-symmetric qubit~\footnote{For this system $\mathcal{P}=\sigma_x$ 
	is the Pauli-$x$ matrix and $\mathcal{T}$ is the complex conjugate operator, $K$: $Kz=z^*$ for $z\in\mathbb{C}$.}:	
	\begin{equation}
	\label{ex1}
%	H_{\text{S}} = \alpha \sigma_{+}\sigma_{-} + \alpha^* \sigma_{-}\sigma_{+} +  \gamma \sigma_{+} +  \gamma^* \sigma_{-},
	H_{\text{S}} = \alpha \sigma_{+}\sigma_{-} + \gamma \sigma_{+} + \text{h.c.}
	=
	\begin{pmatrix}
	 \alpha & \gamma \\
	 \gamma^* & \alpha^*
	\end{pmatrix},
	\end{equation}
	where both $\alpha$ and $\gamma$ can be complex parameters, whereas $\sigma_{+}$ and $\sigma_{-}$ are the raising and lowering fermionic operators.
	This simple model has been extensively studied in the literature~\cite{Deffner2015a,bender_1, bender_2}. Moreover, it has also
	been realized experimentally both in optics~\cite{optic} and semiconductor microcavities~\cite{gao}. We assume the system~(\ref{ex1}) to 
	be coupled to a bosonic heat bath at the inverse temperature $\beta$ via a dephasing interaction. That is to say
	\begin{equation}
		\label{SB}
			H_{\text{B}} = \textstyle{\sum_n} \omega_n a_n^{\dagger}a_n, \quad
			V_{\text{B}} = \textstyle{\sum_n} g_n \left(a_n + a_n^{\dagger} \right),
	\end{equation}
	where $a_n$, $a_n^{\dagger}$ are the bosonic creation and annihilation operator, respectively~\cite{Kumar13}. They obey the canonical 
	commutation relation $[a_n,a_m^{\dagger}]=\delta_{nm}$. The bath's eigenmodes $\omega_n$ and coupling constants $g_n$ are assumed to
	be real. We emphasize that the above bosonic Hamiltonians are hermitian~\cite{Sober15}. Nevertheless, they do \emph{not} commute, \emph{i.e.} 
	$[H_{\text{B}},V_{\text{B}}]\not=0$. This results in nontrivial dynamics and decoherence.
	\begin{figure}[ht]
		\includegraphics[width=.48\textwidth]{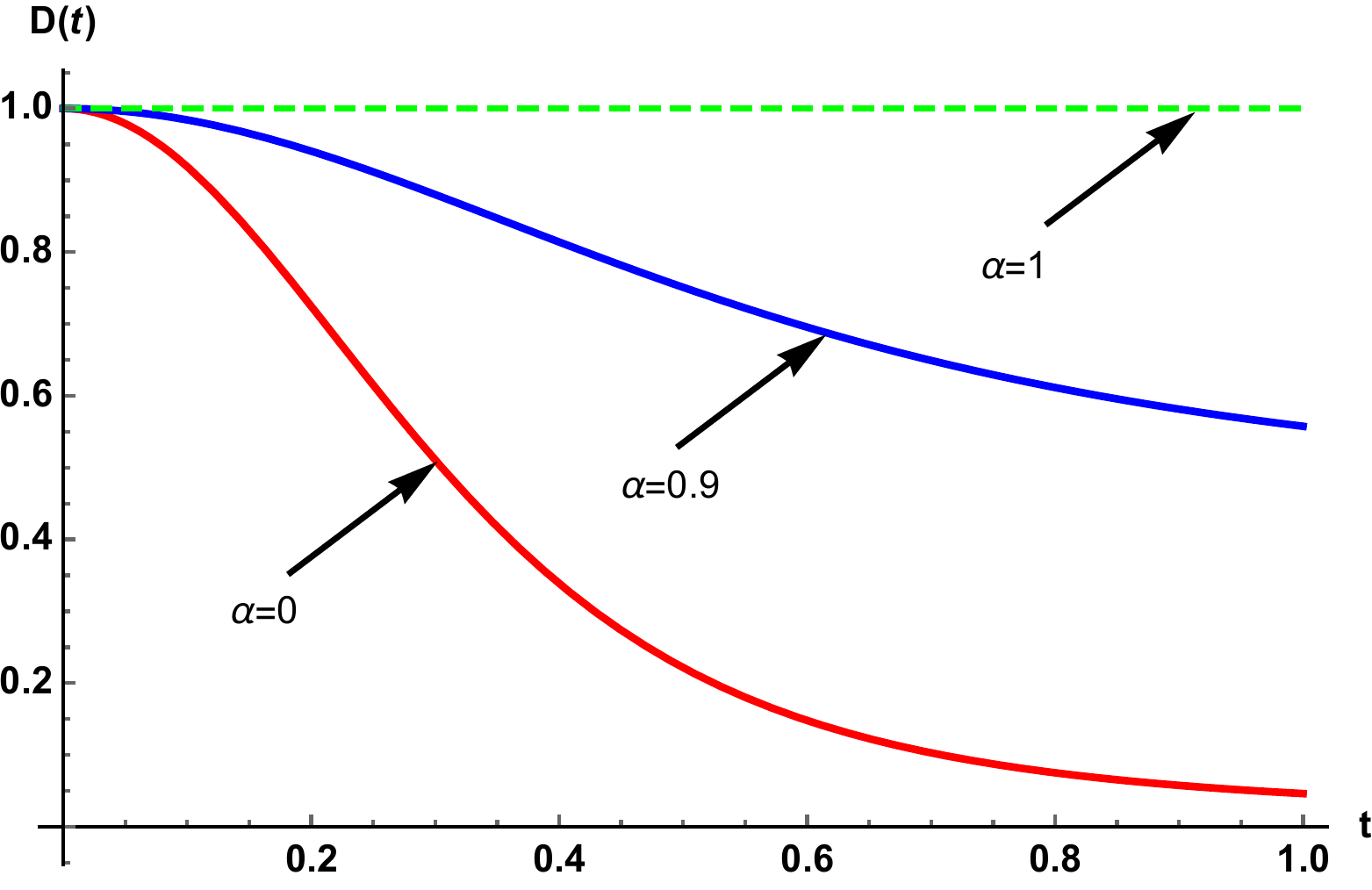}
		\caption{\label{d} (Color online): Critical slowing down of decoherence. The decoherence function $D(t)$ decays more gradually
			as $\alpha$ increases. Eventually, it becomes constant at the critical point $\alpha=1$. Red line corresponds
			to a hermitian system ($\alpha=0$). Parameters are: $\mu=-0.5$, $\,\beta=0.5$ and $J_0=\omega_c=1$.
			For negative $\alpha$ the situation is symmetric [see \emph{e.g.} Eq.~(\ref{simple})].			
			}
	\end{figure}	

	In the following, we explicitly construct the hermitian representation of Hamiltonian~(\ref{ex1}).
	Without any loss of generality, we can choose $\alpha$ to be purely imaginary, \emph{i.e.} $\alpha\rightarrow i\alpha$; we will also set 
	$\gamma=1$. Then, as long as $|\alpha| \leq 1$, the spectrum of $H_{\text{S}}$ is real. It consists of two eigenvalues: 
	$E_{1,2} = \mp \sqrt{1-\alpha^2}$. Simple calculations show that~\cite{Gardas16}
	\begin{equation}
		\label{T}
			T  = 
			U^{\dagger}
			\begin{pmatrix}
				s_1 & 0 \\
								0 & s_2 
			\end{pmatrix}
			U,			
			\quad
			U = 
			\frac{1}{\sqrt{2}}
			\begin{pmatrix}
				i & 1 \\
				-i& 1
			\end{pmatrix},
	\end{equation} 
	where $s_{1,2}=\sqrt{2(1\pm\alpha)}$ and $U$ is unitary.
	Therefore, the corresponding hermitian Hamiltonian reads %$h_{\text{S}} = E_1\sigma_x$, 
	\begin{equation}
	h_{\text{S}} = E_1\sigma_x
	=
	\begin{pmatrix}
	0 & E_1 \\
	E_1 & 0 
	\end{pmatrix},
	\end{equation}
	where $\sigma_x$ is the Pauli-$x$ matrix.
	The resulting model describes the paradigmatic spin-boson system with effective couplings $g_nE_1$~\cite{Fannes88}. 

	In what follows, we assume the initial state of the environment to be the Gibbs state, $\Omega_{\text{B}}=\exp(-\beta H_{\text{B}})/Z$, 
	where $Z=\text{tr}\{\exp(-\beta H_{\text{B}})\}$ is the partition function~\cite{callen}. The reduced dynamics $\varrho(t)=[\varrho_{ij}(t)]_{2\times 2}$ 
	can be obtained exactly~\cite{Gardas11}. Indeed, we have~\cite{Luczka90}
	\begin{equation}
		\label{sol}
			\begin{split}
				\varrho_{11}(t) &= \frac{1}{2} - \Re[\varrho_{12}(0)e^{-iE_1t}]\,D(t), \\
				\varrho_{12}(t) &= \varrho_{11}(0) - \frac{1}{2} + i\Im[\varrho_{12}(0)e^{-iE_1t}]\,D(t), %\\
				%\varrho_{21}(t) &= \varrho_{12}(t)^* \quad \text{and} \quad \varrho_{22}(t) = 1-\varrho_{11}(t).
			\end{split}	
	\end{equation} 
	Moreover, $\varrho_{22}(t)=1-\varrho_{11}(t)$ and $\varrho_{21}(t)=\varrho_{12}(t)^*$~\footnote{Note, $\unexpanded{\langle 
	H_{\text{B}}\rangle}=2\varrho_{11}(0)-1$ is constant as one would expect.}.
	Above, symbols $\Re(z)$ and $\Im(z)$ denote the imaginary and real parts of a complex number $z$, respectively,
	whereas the decoherence function $D(t)=\exp(-E_1^2\gamma(t))$ quantifies decoherence~\cite{Plenio14}. Information
	regarding the environment is encoded in the temperature-dependent function $\gamma(t)$,
	\begin{equation}
		\label{gamma}
			\gamma(t) = \int_0^{\infty}\,d\omega\frac{J(\omega)}{\omega^2}\,\left(1-\cos\omega t\right)\coth\left(\frac{\hbar\beta\omega}{2}\right),
	\end{equation} 
	where $J(\omega)=\sum_n |g_n|^2\delta(\omega-\omega_n)$ is the spectral density that characterizes the environment.
	Typical examples include $J(\omega) = J_0\omega^{1+\mu}\exp(-\omega/\omega_c)$ for some predefined constants $J_0$, $\mu$ and 
	$\omega_c$~\cite{Leggett87}. For example, when $\mu=0$ (Ohmic case) and $\beta\omega_c \gg 1$, in the long time limit the decoherence 
	function behaves like (exponential relaxation~\cite{Luczka90}) 
	\begin{equation}
	\label{simple}
		D(t) \sim \exp\left[-\pi J_0\left(1-\alpha^2\right)t/\beta\right].
	\end{equation}
	The reduced dynamics~(\ref{sol}) can also be expressed using the Kraus representation directly~\cite{Luczka90}.	
	The reduced dynamics for the original $\mathcal{P}\mathcal{T}$ - symmetric qubit~(\ref{ex1}) can now be calculated
	as $\rho(t)=T^{-1}\varrho(t)T$, where $T$ is given by Eq.~(\ref{T}).
		
	Since $D(t)\rightarrow 0$, from Eq.~(\ref{sol}) it is evident that the environment will eventually destroy the coherent dynamics of the 
	system. However, this process can be controlled by changing $\alpha$ [cf. Eq.~(\ref{simple})]. Indeed, as depicted in Fig.~\ref{d}, 
	decoherence becomes slower [\emph{i.e.} $D(t)$ decays more gradually] as $\alpha$ increases. Moreover, when $\alpha\rightarrow 1$ the 
	decoherence process becomes suppressed completely~\cite{Viola98}. However, when the system is hermitian, \emph{i.e.} $\alpha\rightarrow 0$, decoherence becomes severe and quickly destroys any coherence.
	
	To explain this phenomenon we notice that when $\alpha>1$ all eigenvalues of $H_{\text{S}}$ are complex. Therefore, $\alpha=1$ can be seen
	as a critical point separating two physically distinct regimes. As $E_1 \rightarrow 0$ when $\alpha\rightarrow 1$, it takes longer for the
	system to complete one oscillation (in the Hilbert space) in close proximity of the critical point. Precisely at that point the dynamics 
	``freezes out'' completely. 
	This critical slowing down also affects decoherence (\emph{critical slowing down of decoherence}) because of the effective coupling strengths 
	$g_nE_1$ that also depend on $\alpha$. Setting $g_n=1/E_1$ removes the $\alpha$-dependence from the interaction and assists 
	decoherence~\cite{Sinayskiy12}. At the critical point $\rho(t)\rightarrow \infty$; however, $\langle O_{\text{S}}(t)\rangle=\tr{\varrho(0)o_{\text{S}}}$ 
	is finite. Therefore, when $\alpha \rightarrow 1$ expectation values are determined only by the initial condition and remain unchanged. This
	is due to $h_{\text{S}} \rightarrow 0$ -- the ``freezing out'' of the dynamics.
    
    A similar dynamical behavior manifesting itself through the ``freezing out'' scenario has already been observed in closed quantum systems. 
    The 1D Ising model~\cite{Dziarmaga05}, where the Kibble-Zurek mechanism~\cite{Kibble76,Zurek85} can be applied, provides one example,
    another example is the Landau-Zener problem~\cite{Quintana13} of a two level quantum system that supports the Kibble-Zurek mechanism~\cite{Damski05}. 
	\paragraph{Summary.}		 
	We have investigated a $\mathcal{P}\mathcal{T}$-symmetric quantum system coupled to an external environment. To this end, we have considered
	a particular scenario where there is no heat exchange between these two systems and only quantum information is allowed to enter/leave the 
	system. This phenomenon is known as pure decoherence or dephasing. We have shown how to derive the reduced dynamics using a quantum canonical 
	transformation.
	
	Moreover, we have studied an experimentally relevant example, namely a $\mathcal{P}\mathcal{T}$-symmetric qubit. Such a system can be 
	realized \emph{e.g.} in optics~\cite{optic} and microcavities~\cite{gao}. In contrast to hermitian qubits, this system exhibits a phenomenon 
	that we identified as \emph{critical slowing down of decoherence}. As we have argued, such behavior is characteristic of every open 
    $\mathcal{P}\mathcal{T}$-symmetric and any quantum system whose spectrum can be divided in two physically different regimes~\cite{Peng14}. 
    When a system approaches the critical point separating these two regimes its dynamics ``freezes out''. This critical slowing down also affects 
    decoherence due to the dephasing interaction. Concluding, this behavior suggests that $\mathcal{P}\mathcal{T}$-symmetric qubits may be more 
    robust against decoherence and therefore be better suited as components in quantum computers~\cite{Johnson11,Lanting14,Raussendorf01}.

    Experimental setups that are sensitive enough to detect the $\mathcal{P}\mathcal{T}$-symmetry breaking (\emph{i.e.} the critical point) 
    should also be able to capture the critical slowing-down~\cite{Peng14}. Thus, \emph{critical slowing-down of decoherence} can be testable
    as well, provided there is no heat exchanged with the environment. Such induced dephasing, however, can be realized experimentally~\cite{Alvarez10}. 
	
	\paragraph{Acknowledgments.} 
	 We thank Wojciech H. Zurek for stimulating discussions. This work was supported by the Polish Ministry of
	 Science and Higher Education under project Mobility Plus 1060/MOB/2013/0 (B.G.); S.D. acknowledges financial support from the
	 U.S. Department of Energy through a LANL Director's Funded Fellowship.		
%
%\bibliography{PT}
\end{document}